\documentclass[12pt]{elsarticle}

\usepackage{amssymb}
\usepackage{amsmath}
\usepackage{subcaption}
\usepackage{multirow}
\usepackage[table]{xcolor}
\usepackage{url}
\usepackage{geometry}
\usepackage{booktabs}

\journal{International Journal of Electrical Power \& Energy Systems}

\begin{document}

\begin{frontmatter}



\title{Impact of Communication Delay and Sampling on Small-Signal Stability of IBR-rich Power Systems}

\author[wsu]{Saugat Ghimire, Vaithianathan "Mani" Venkatasubramanian} 
\author[rte]{Gilles Torresan} 

\affiliation[wsu]{organization={Washington State University},
            addressline={}, 
            city={Pullman},
            postcode={99164}, 
            state={WA},
            country={USA}}
            
\affiliation[rte]{organization={RTE},
            addressline={}, 
            city={Paris},
            postcode={92073}, 
            state={},
            country={France}}

\begin{abstract}
The growing adoption of inverter-based resources (IBRs) has introduced unprecedented dynamics in power systems, resulting in oscillations across a broad spectrum of frequencies. Communication delay between the plant-level control and the inverter-level control in IBR plants has been recognized as one of the causes of such oscillations and a factor that impacts the system's stability. The control signals from the plant-level controller also experience sampling, with the sampled values held constant by the hold elements for the duration of the sampling period. This also has a bearing on the response of IBR plants. In this paper, we analyze the impacts of communication delay and sampling of control signals between plant-level control and inverter-level control of grid-following IBR plants on the small-signal stability of power systems. The underlying fundamentals of communication delay and sampling are revisited to explain the observed responses. Our findings emphasize the unique effects of communication delay and sampling period on the stability of IBR-rich power systems and suggest strategies to mitigate their detrimental impacts. The work also highlights the need for more accurate approaches for small-signal stability analysis of such systems.
\end{abstract}




\end{frontmatter}


\section{Introduction}
\label{sec:Introduction}

With the increasing penetration of IBR plants connected to bulk power systems (BPSs), power networks worldwide are experiencing unforeseen dynamics. Undesirable behaviors such as oscillations ranging from low to super-synchronous frequencies \cite{RealSSOs, solar2017deploying} have raised concerns about the stability of modern power systems. Increased complexity and faster dynamics accompanying the IBR plants demand fresh considerations in the stability analysis of IBR-rich power networks. A significant amount of work can be found in the literature that provides insights and explanations to various IBR-related oscillations observed in the real systems \cite{RealSSOs, solar2017deploying, NERC2017, fan2017explanation, ERCOTOsc2012, fan2022SolarOsc, Chetan}. Communication delay between the plant-level and the inverter-level controllers present in IBR plants has been identified as one of the causes of such oscillations \cite{RealSSOs, fan2022SolarOsc, OpChallangesFan, Chetan}. 

An analysis of 4 Hz oscillations observed in a wind power plant in Texas is presented in \cite{RealSSOs}, and communication delay is recognized as a critical factor for explaining the observed behavior. Similarly, in \cite{fan2022SolarOsc}, modeling of communication delay is deemed necessary to analyze the 0.1 Hz oscillations detected in a BPS-connected solar PV power plant. Recognizing the effect of communication delays, some approaches have also been proposed in the literature to design controllers to alleviate its detrimental impact \cite{ramasubramanian2022differentiating, FanDesignConsideration}. 

In addition to suffering from communication delays, the control signals from plant-level control to inverters are also sampled at regular intervals. The sampling period of the sampler impacts the system's stability. However, most of the works found in the literature do not consider the effect of this sampling, and those that do, model the effect of the sampling period by approximating it as a constant delay \cite{OpChallangesFan, ramasubramanian2022differentiating}. No previous study has specifically examined the independent impact of communication delays and sampling in IBR plants on the power system's small-signal stability. Some works have considered the different effects of communication delay and sampling period on synchronous generator controls, but the assumptions in these studies are not always valid for IBR-rich systems with multiple independent delays and samplers \cite{Milano2019, Hiskens2020}. The communication delay and the sampling period in IBR plants are generally substantial, typically around 500 milliseconds, and can extend up to 10 seconds, as noted in \cite{Chetan}. Table \ref{tab:RTEdata} shows these values for some IBR plants in the RTE system. Proper modeling of such considerable delays and sampling periods is vital for stability analysis, particularly for systems involving fast dynamics. The small signal stability analysis of time-delayed systems is complicated as the characteristic equation of such systems has infinite roots. Thus, all the poles of the systems cannot be calculated. Approximations such as a first-order low-pass filter \cite{WECCPV}, the Pade approximation \cite{vemula2023impact, chowPade}, or finite-element-based approximations \cite{Milano2012, wu2019delay} are used to obtain approximate roots of the systems.  
Such approximations are effective for small delays and sampling periods but can fail for values common in IBR plants \cite{Milano2019}. As demonstrated in this work, greater caution should be exercised while employing these approximations in analyzing IBR-rich power systems with changing and faster dynamics.

\begin{table}[h]
\footnotesize
\centering
\caption{Communication delays and sampling periods of IBR plants}
\label{tab:RTEdata}
\begin{tabular}{@{}ccc@{}}
\toprule
\textbf{Type} & \textbf{Sampling Time (ms)} & \textbf{Comm. Delay (ms)} \\ \midrule
PV   & 500                & 100       \\ 
PV   & 1000               & 750       \\
PV   & 60                 & 118       \\       
Wind & 300                & 50      \\        
Wind & 150                & 9      \\          \bottomrule
\end{tabular}
\end{table}

In this paper, we separately investigate the effect of communication delays between plant-level and inverter-level controls, as well as the impact of sampling of plant-level control signals in grid-following (GFL) power plants, on the small-signal stability of the power system. Extensive simulations and fundamental analysis provide insight into the dynamic behavior of such systems. The paper employs two well-tested ring-down analysis algorithms, namely, the Matrix Pencil (MP) and Eigensystem Realization Algorithm (ERA) \cite{ManiEventAlgo, RTEimplementation}, for modal estimation to analyze the changes in frequency and damping of the dominant modes in various test cases. While most previous studies that have considered time delay primarily focus on estimating the delay margin, the critical delay value beyond which the system becomes unstable, this work demonstrates that a sampled time-delayed system can exhibit multiple distinct regions of stability, interspersed with regions of instability. The paper presents a theoretical analysis of such responses of delayed and sampled systems in simple test cases, first to derive analytical insight into the stability phenomena. The key observations from the study are then shown to closely align with the responses observed in IBR-rich power system examples.

The key contributions of the paper can be summarized as follows.
\begin{itemize}\vspace{-0.1in}
    \item Analysis of the {\em significant and distinct impacts} of communication delay and sampling on small-signal stability of power systems with GFL plants. \vspace{-0.1in}
    \item Identification of limitations of the prevalent small-signal stability analysis methods for IBR-rich systems with large communication delays and sampling periods. {\vspace{-0.1in}}
    \item Demonstration of counterintuitive examples showcasing multiple regions of stability, and stability enhancement by slowing down the sampling. \vspace{-0.1in}
    \item Recommendations on control strategies to alleviate the destabilizing effects of communication delays and sampling.
\end{itemize}

The rest of the paper is organized as follows. Section \ref{sec:Architecture} discusses a general control architecture of grid-following power plants. Section \ref{sec:Analysis} analyzes the impact of delay and sampling on the stability of dynamic systems and highlights their differences. Section \ref{sec:SMIB} evaluates the effects of communication delay and sampling of commands using different simulation scenarios in a single-machine infinite bus (SMIB) system, while a similar analysis for the two-area Kundur system is provided in Section \ref{sec:Kundur}. Control strategies to reduce the detrimental effects of communication delay and sampling periods are discussed in Section \ref{sec:Mitigation}. Finally, Section \ref{sec:Conclusion} concludes the paper. 
\vspace{-0.1in}

\section{Control Architecture of IBR Plants}
\label{sec:Architecture}

\begin{figure*}[htbp]
\centerline{\includegraphics[width=0.98\textwidth]{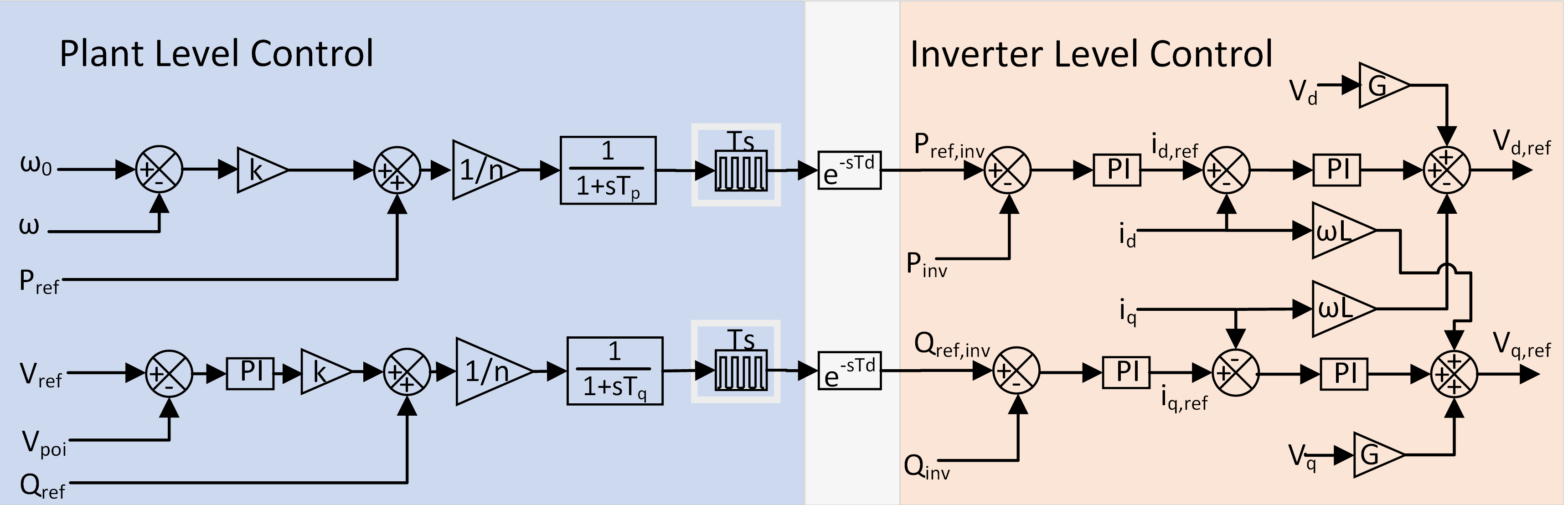}}
\caption{Control architecture of a grid-following power plant.}
\label{fig:ControlArchitecture}
\vspace{-0.1in}
\end{figure*}

A BPS-connected IBR power plant comprises several inverter devices connected to power sources, such as solar photovoltaic (PV) panels, wind generators, or battery energy storage systems (BESS). The inverters are centrally coordinated by a plant-level controller, which regulates the generation of active and reactive power according to set values. The plant-level controller is housed within the plant's control center, which can be located at a significant distance from individual inverter devices, especially in offshore and deep-sea wind power plants (WPPs). In such cases or for plants with satellite communications, the transfer of control signals between these controllers can be hindered by pronounced communication delays. Such communication delays impact the small-signal stability of the system \cite{pan2024primary}.

A typical control architecture of the grid-following control of a BPS-connected IBR power plant is presented in Fig. \ref{fig:ControlArchitecture}. The plant level control has active and reactive power droops and voltage control. It is responsible for maintaining the set power flow and the voltage at the point of interconnection (POI). The plant-level control sends the active and reactive power commands to the individual inverter-level control. The commands from the plant-level control are sampled at an interval $T_s$ before being transferred to the inverters, and there is a communication delay of $T_d$ seconds, represented by $e^{-sT_d}$, for the sampled commands to reach the inverters. The inverter level control has power control loops and inner current control loops, which regulate the power output of the inverters.

\section{Stability Concerns from Delay and Sampling}
\label{sec:Analysis}

The controllers of GFL plants incorporate sampling and communication delays as integral components. In this section, we analyze the effects that delays and sampling have on the stability of dynamic systems.

\subsection{Effect of delay on stability of a system}
\label{sec:Analysis:Delay}

\begin{figure}[htbp]
\centering
\begin{subfigure}[b]{0.4\linewidth}
\includegraphics[width=\linewidth]{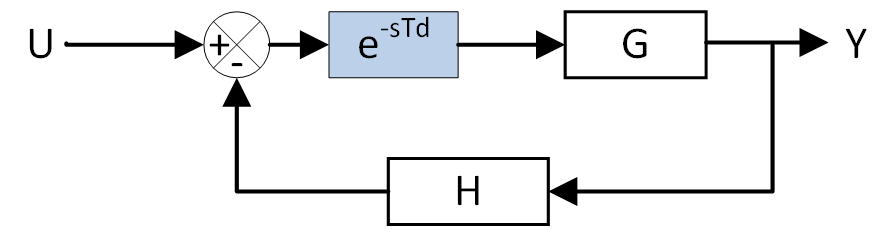}
\caption{Time delay in the forward path}
\label{fig:SimpleDelayedSystem_1}
\end{subfigure}
\begin{subfigure}[b]{0.4\linewidth}
\includegraphics[width=\linewidth]{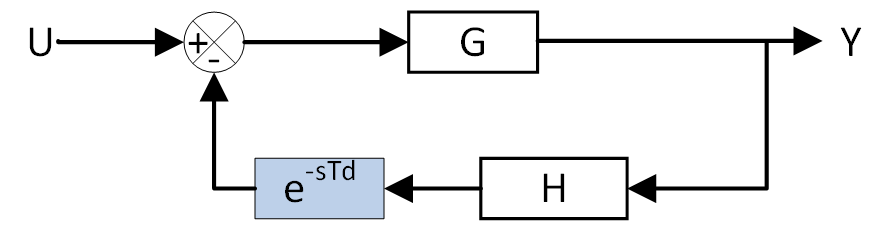}
\caption{Time delay in the feedback path}
\label{fig:SimpleDelayedSystem_2}
\end{subfigure}
\caption{Closed-loop time-delayed systems.}
\label{fig:SimpleDelayedSystem}
\end{figure}

An ideal delay does not affect the magnitude response of an open-loop system, so the impact of internal delay, $T_d$, in the system shown in Fig. \ref{fig:SimpleDelayedSystem_1}, can be analyzed using the phase response of the open-loop system G(s)H(s).  Given $\angle GH(j\omega)$ as the phase response of G(s)H(s), the open-loop phase response of the delayed system in Fig. \ref{fig:SimpleDelayedSystem_1} is obtained as: 

\vspace{-0.05in}
\begin{equation}
\label{eq: delay on phase}
\angle (e^{-j\omega T_d}GH(j\omega)) = \angle GH(j\omega) - \omega T_d
\end{equation}

Equation (\ref{eq: delay on phase}) implies that the delay changes the phase crossover frequencies of a system. At phase crossover frequencies, the 180° phase shift causes negative feedback to act as positive feedback. An open-loop gain exceeding unity at these points results in amplification and leads to instability. Thus delay can affect the closed-loop stability of the system by shifting phase crossover frequencies, and thereby, its gain margin. Higher delays also result in closely spaced phase crossover points, increasing the probability of negative gain margin and instability. (\ref{eq: delay on phase}) also implies that the frequency responses for the two systems shown in Fig. \ref{fig:SimpleDelayedSystem} are the same, and the same stability analysis holds for both systems. 

\begin{figure}
\centering
\begin{subfigure}[b]{0.24\linewidth}
\includegraphics[width=\linewidth]{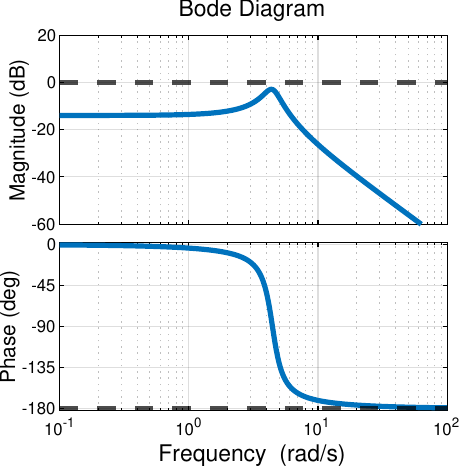}
\caption{}
\label{fig:Bode_delay0.2}
\end{subfigure}
\begin{subfigure}[b]{0.24\linewidth}
\includegraphics[width=\linewidth]{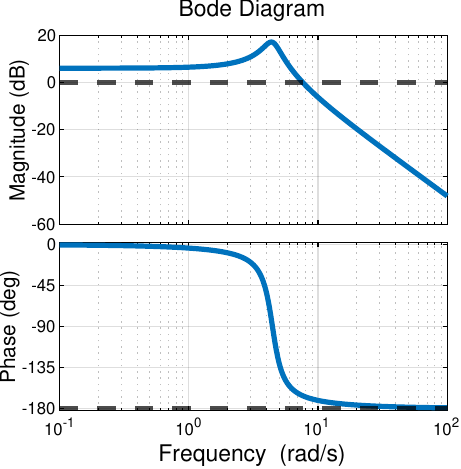}
\caption{}
\label{fig:Bode_delay2}
\end{subfigure}
\begin{subfigure}[b]{0.24\linewidth}
\includegraphics[width=\linewidth]{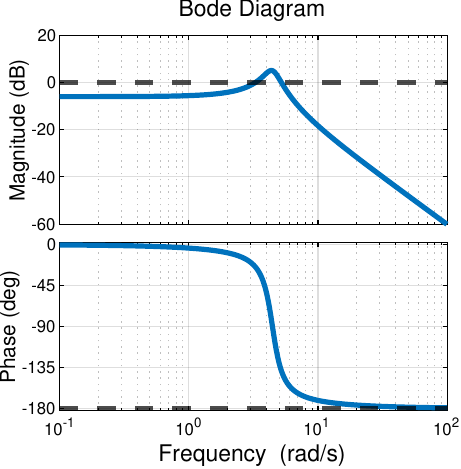}
\caption{}
\label{fig:Bode_delay0.5}
\end{subfigure}
\vspace{-0.05in}
\caption{Frequency response of a system with (a) no gain crossover, (b) one gain crossover, (c) two gain crossover frequencies.}
\label{fig:BodeDelay}
\vspace{-0.2in}
\end{figure}

Based on the properties of G(s)H(s), the open-loop system can have either no gain crossover (i.e., no crossing of 0 dB line), one gain crossover, or multiple gain crossovers as shown in Fig. \ref{fig:BodeDelay}. When there is no gain crossover, there is no phase or delay margin for this kind of system. The closed-loop system's damping changes with a change in the delay because of a change in the gain margin, but the system never becomes unstable for any delay value as the gain margin is always positive. A system with one gain crossover has a delay margin given by the ratio of the phase margin to the gain crossover frequency. The closed-loop system becomes unstable for delay values larger than the delay margin. And, systems with multiple gain crossovers have multiple delay margins. The closed-loop system becomes unstable for some delay values larger than the smallest delay margin. However, stability regions can also exist beyond the delay margin.

\begin{figure}[htbp]
  \centering
  \begin{minipage}[b]{0.4\textwidth}
  \centering
    \includegraphics[width=\textwidth]{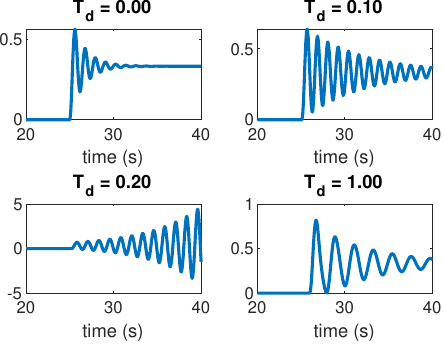}
    \caption{Step responses of the system with varying Td.}
    \label{fig:DelayStepResponses}
  \end{minipage}
  \hspace{0.1in}
  \begin{minipage}[b]{0.4\textwidth}
  \centering
    \includegraphics[width=0.88\textwidth]{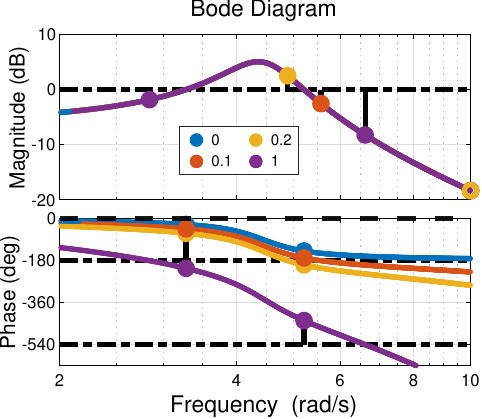}
    \caption{Change in gain margin with varying Td.}
    \label{fig:BodeDelayVaried}
  \end{minipage}  
  \vspace{-0.1in}
\end{figure}

The step responses of an example system with two gain crossover frequencies are presented in Fig. \ref{fig:DelayStepResponses} for different delay values. For this system, delayed system A, G(s) and H(s) are defined as:

\begin{equation}
         G(s) = \frac{k\omega^2_n}{s^2 + 2\zeta\omega_n + \omega^2_n}  \ \ \mbox{and} \ \ H(s) = 1   \label{eq:SimpleSystem} 
\end{equation}

with $k=0.5$, $\omega_n = 4.44$ and $\zeta = 0.1414$. The frequency and damping ratio (DR) estimated using the ring-down algorithms are reported in Table \ref{tab:DelayModes}. It can be observed that the system's damping initially decreases with an increase in delay, eventually making it unstable for a delay of 0.2 seconds. The system returns to stability when the delay value is increased to 1 second. Such behavior can be described by the open-loop frequency response of this system, as presented in Fig. \ref{fig:BodeDelayVaried}. As seen, the positive gain margin initially decreases, becomes negative for some values of delays, and then returns to positive values on further increasing the delay, bringing the system back to stability. Moreover, the observed leftward shift of the phase crossover frequency with increasing delay accounts for the reduction in the mode frequency from 0.86 Hz to 0.46 Hz.

\begin{table}
    \centering
    \footnotesize
    \caption{Model Estimates for Varying $T_d$.}
    \label{tab:DelayModes}
    \begin{tabular}{@{}ccc@{}}
        \toprule
         \textbf{$T_d$ (s)} & \textbf{Freq. (Hz)} & \textbf{DR (\%)} \\
         \midrule
         0   & 0.86 & 11.54 \\
         0.1 & 0.85 & 2.83 \\
         0.2 & 0.80 & -3.45 \\
         1   & 0.46 & 5.26 \\
         \bottomrule
    \end{tabular}
\end{table}

\subsection{Effect of sampling on stability}
\label{sec:Analysis:Sampling}

\begin{figure}[htbp]
\centerline{\includegraphics[width=0.4\textwidth]{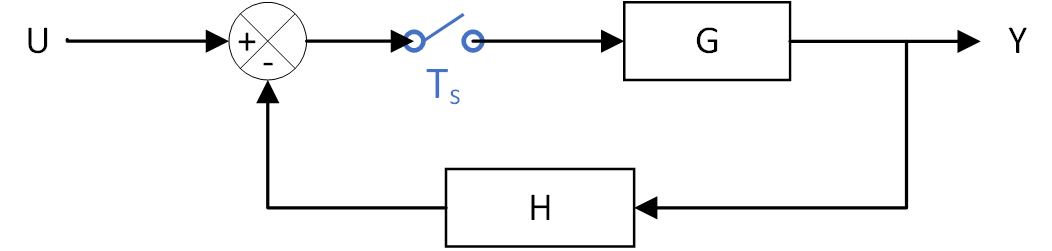}}
\caption{Closed-loop sampled test system.}
\label{fig:SimpleSampledSystem_1}
\end{figure}

The behavior of sampled systems can be analyzed using their discrete representation. In order to analyze the effect of sampling period, consider the sampled system shown in Fig. \ref{fig:SimpleSampledSystem_1}. The sampling period is $T_s$, and G represents the plant transfer function with the zero-order hold element. 

The discrete transfer function of the system can be written as:
\begin{equation}
\label{eq:SingleRateDiscrete}
        \frac{Y(z)}{U(z)} = \frac{G(z)}{1+GH(z)}
\end{equation}

\noindent where, $z = e^{-sT_s}$.

Considering the same G(s) and H(s) defined by (\ref{eq:SimpleSystem}) and assuming the poles of G(s) to be $s = -\alpha \pm j\beta$, the z transform of $G(s)$ with zero-order hold is obtained as:
\begin{equation}
    \label{eq:SampledSystemz}
    G(z) = \frac{Az+B}{z^2 - 2e^{-\alpha T_s}cos(\beta T_s)z + e^{-2\alpha T_s}}
\end{equation}
where, 

$A = k - \frac{k\omega_n}{\beta}e^{-\alpha T_s}sin(\beta T_s + tan^{-1}(\frac{\beta}{\alpha}))$ ,
$B = ke^{-2\alpha T_s} +\frac{k\omega_n}{\beta}e^{-\alpha T_s}sin(\beta T_s - tan^{-1}(\frac{\beta}{\alpha}))$
\vspace{0.01in}

The characteristic equation of the closed-loop system in Fig. \ref{fig:SimpleSampledSystem_1} is:
\vspace{-0.1in}
\begin{equation}
\label{eq:SampledCharEq}
 z^2 + (A-2e^{-\alpha T_s}cos(\beta T_s))z+(B+e^{-2\alpha T_s}) = 0    
\end{equation}

\begin{figure}[htbp]
  \centering
  \begin{minipage}[b]{0.4\textwidth}
    \includegraphics[width=\textwidth]{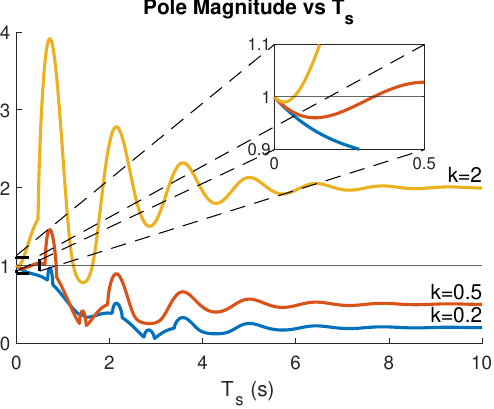}
    \caption{Change of pole magnitude of the sampled system with varying Ts.}
    \label{fig:PoleMagSRD}
  \end{minipage}
  \hspace{0.1in}
  \begin{minipage}[b]{0.4\textwidth}
    \includegraphics[width=\textwidth]{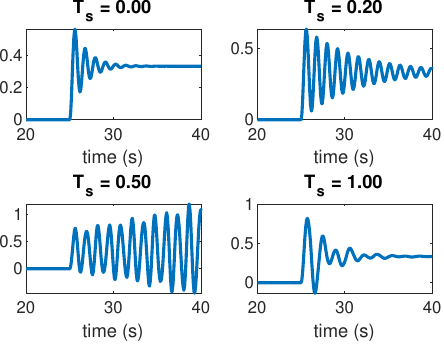}
    \caption{Step responses of the system with varying Ts.}
    \label{fig:SampleStepResponses}
  \end{minipage}
\end{figure}

Equation (\ref{eq:SampledCharEq}) is quadratic in z with coefficients that are functions of $T_s$. The presence of sinusoidal terms in the coefficients implies periodicity in the nature of the system's poles. Fig. \ref{fig:PoleMagSRD} shows the magnitude of the pole of the systems with different values of k as the sampling period varies from 0 to 10 seconds. The system is unstable when the magnitude of the z-domain pole is greater than 1. For k = 0.2, the system is stable for all values of $T_s$. For k = 0.5, it is evident that the system has a small region of instability between $T_s$ = 0.33 and 0.85 seconds and is stable for all other values of sampling periods. For k = 2, the system has two regions of stability, first with very small Ts, from 0 to 0.06 seconds, and the second with larger Ts, from 1.22 to 1.63 seconds. Further, equation (5) shows that as $T_s \to \infty, |z| \to k$, implying stability at large sample periods for systems with open-loop dc gain ($k$) less than unity, and instability otherwise.

The step responses of the sampled system with k = 0.5 are shown in Fig. \ref{fig:SampleStepResponses} for different sampling periods, and Table \ref{tab:SamplerModes} compares the frequency and damping ratio of the system's poles obtained using the discussed discrete-time analysis,  Pade approximation with order 10, and the ring-down algorithms, MP and ERA. The results show that with the increase in the sampling period, the damping of the system mode initially decreases to the point of instability and then returns to stability. 

Table \ref{tab:SamplerModes} further shows that the results of discrete-time analysis match reasonably with the mean estimates of MP and ERA algorithms. However, the results obtained with Pade approximations differ significantly, even with a high order of 10. Additionally, Pade approximation incorrectly identifies the system as unstable for $T_s = 0.2$ seconds while its stability can be observed in the simulation result in Fig. \ref{fig:SampleStepResponses} and with the proven measurement-based algorithms. This demonstrates the inappropriateness of modeling the effect of sampling as a constant delay in power systems, even for systems with slow dynamics, i.e., less than 1 Hz. 

\begin{table}
\footnotesize
\centering
\caption{Comparison of Mode Estimates Obtained from Different Methods}
\label{tab:SamplerModes}
\begin{tabular}{@{}ccccccc@{}}
\toprule
& \multicolumn{2}{c}{\textbf{Discrete}}  & \multicolumn{2}{c}{\textbf{Pade}}  & \multicolumn{2}{c}{\textbf{MP/ERA}}   \\ \cline{2-7} 
\multirow{-2}{*}{\textbf{\begin{tabular}[c]{@{}c@{}}Ts\\ (s)\end{tabular}}} & \multicolumn{1}{c}{\textbf{\begin{tabular}[c]{@{}c@{}}Freq\\ (Hz)\end{tabular}}} & \textbf{\begin{tabular}[c]{@{}c@{}}DR\\ (\%)\end{tabular}} & \multicolumn{1}{c}{\textbf{\begin{tabular}[c]{@{}c@{}}Freq\\ (Hz)\end{tabular}}} & \textbf{\begin{tabular}[c]{@{}c@{}}DR\\ (\%)\end{tabular}} & \multicolumn{1}{c}{\textbf{\begin{tabular}[c]{@{}c@{}}Freq\\ (Hz)\end{tabular}}} & \textbf{\begin{tabular}[c]{@{}c@{}}DR\\ (\%)\end{tabular}} \\ 
\midrule
0 & \multicolumn{1}{c}{0.86} & 11.54 & \multicolumn{1}{c}{0.86} & 11.54 & \multicolumn{1}{c}{0.86} & 11.54 \\
\hline
0.2 & \multicolumn{1}{c}{0.84} & 3.23 & \multicolumn{1}{c}{0.79} & -3.45 & \multicolumn{1}{c}{0.84} & 3.2 \\ 
\hline
0.5 & \multicolumn{1}{c}{0.77} & -1.15 & \multicolumn{1}{c}{0.64} & -7.54 & \multicolumn{1}{c}{0.77} & -1.15 \\ 
\hline
\cellcolor[HTML]{FFFFFF} & \multicolumn{1}{c}{\cellcolor[HTML]{FFFFFF}0.38} & 15.80 & \multicolumn{1}{c}{\cellcolor[HTML]{FFFFFF}0.46} & 5.26 & \multicolumn{1}{c}{0.37} & 15.95 \\ 
\rowcolor[HTML]{FFFFFF} 
\multirow{-2}{*}{\cellcolor[HTML]{FFFFFF}1} & \multicolumn{1}{c}{\cellcolor[HTML]{FFFFFF}-} & - & \multicolumn{1}{c}{\cellcolor[HTML]{FFFFFF}1} & 11 & \multicolumn{1}{c}{\cellcolor[HTML]{FFFFFF}0.63} & 9.42 \\ 
\bottomrule
\end{tabular}
\vspace{-0.1in}
\end{table} 

The discrete analysis method provides good results for analyzing the stability of the sampled system. However, it is well known that such analysis can be used to study systems with the sampling frequency faster than the system's dynamics. The poles obtained from such analysis can only represent the modes of the system with frequencies less than $1/(2T_s)$ Hz, as dictated by the Nyquist Theorem. Consequently, such system representation results in high discretization error and loss of information on the system's fast dynamics when the sampling period is large. This can be observed for $T_s= 1$ second in Table \ref{tab:SamplerModes}. In this case, the system has two modes, one at 0.37 Hz and the other at 0.63 Hz, as obtained with ring-down algorithms and visible in the step response. However, the discrete analysis only captures the 0.38 Hz mode, i.e., the one less than 0.5 Hz, and misses the dynamics of the 0.63 Hz mode. Although suitable for determining the stability of single-rate single-loop systems with a large sampling period, such discrete analysis cannot be performed to study systems with multiple samplers operating at different frequencies. Hence, this method of discrete representation is unsuitable for the analysis of the IBR-dominated power systems.

\vspace{-0.05in}
\section{Effect of Communication Delay and Sampling Time on an SMIB System}
\label{sec:SMIB}

\begin{figure}[htbp]
\centerline{\includegraphics[width=0.5\textwidth]{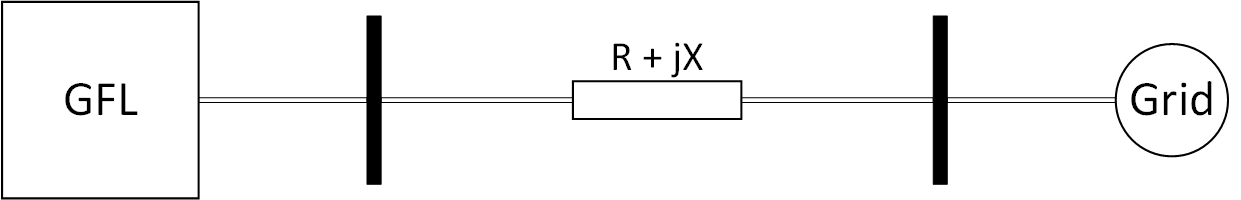}}
\caption{A GFL plant connected to a grid.}
\label{fig:SMIB}
\vspace{-0.1in}
\end{figure}

In this section, we analyze the effects of the sampling period and communication delay on the stability of a power system consisting of a GFL plant connected to an infinite bus. The plant is connected to the infinite bus via a transmission line represented by a short-line model. Simulations are carried out in PSCAD \cite{manitoba1994pscad} for different scenarios, and the dominant modes of the system are estimated using MP and ERA on the PSCAD measurements.

\begin{figure}[htbp]
    \centering
    \begin{minipage}[t]{0.48\textwidth}
        \centering
        \scriptsize
        \captionof{table}{Main Parameters of GFL Plant.}
        \label{tab:GFLPara}
        \begin{tabular}{@{}c@{}c@{}}
            \toprule
            \textbf{Parameter}                        & \textbf{Value (pu)} \\ \midrule
            Filter Inductance                         & 0.009       \\ 
            Filter Resistance                         & 0.016       \\ 
            Filter Capacitance                        & 2.55        \\ 
            Coupling inductance                       & 0.002       \\ 
            Coupling resistance                       & 0.003       \\ 
            Plant-level P-f droop                     & 0.05        \\ 
            Plant-level Q-V droop                     & 0.05           \\ 
            Plan-level voltage control PI             & 0.003 + 0.09/s \\ 
            Plant-level P/Q filter time constant
            & 0.02 sec\\
            P/Q control PI                            & 2 + 20/s      \\ 
            Current control PI                        & 0.38 + 0.7/s  \\ 
            Current control feed-forward gain         & 1.0           \\ 
            Power measurement filter cutoff frequency & 0.132         \\ 
            PLL filter cutoff frequency               & 1.32          \\ 
            PLL PI                                    & 50 + 410/s     \\ \bottomrule
        \end{tabular}
    \end{minipage}%
    \hfill%
    \begin{minipage}[t]{0.47\textwidth}
    \vspace{-0.1in}
        \centering
        \label{}
        \includegraphics[width=\textwidth]{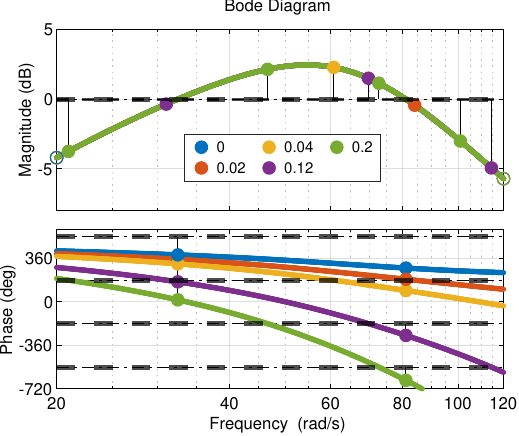}
        \caption{Frequency response of the SMIB active power loop with varying $T_d$.}
        \label{fig:SMIB_Bode}
    \end{minipage}%
\end{figure}

The power plant is rated at 900 MVA, and the system is operated at 20 KV. The details of the studied GFL model can be found in \cite{kenyonGFL}, and the main parameters are listed in Table \ref{tab:GFLPara}. The plant is set to generate 800 MW of active power and 150 MVAR of reactive power. For the base case, a short circuit ratio (SCR) of 3.5, which aligns with the recommended values in the RTE\footnote{Criterion §3.6 in \cite{rte2024}  specifies "b" the maximum impedance for connection to the RTE network as 0.3 pu for IBR plants rated 50–250 MVA, 0.54 pu for 250–800 MVA, and 0.6 pu for \>800 MVA, which correspond to minimum SCR values of 3.33, 1.85, and 1.67, respectively.} and Australian \cite{fluenceenergy2024} systems, is considered at the point of connection. A value of 10 for the inductive reactance-to-resistance (X/R) ratio is adopted. The system's responses under different conditions are analyzed by applying a disturbance in the form of a pulse with a magnitude of 0.01 pu and a duration of 0.01 seconds to the active power set point of the GFL inverter. 

\begin{figure}[htbp]
\centering
  \includegraphics[width=0.8\textwidth]{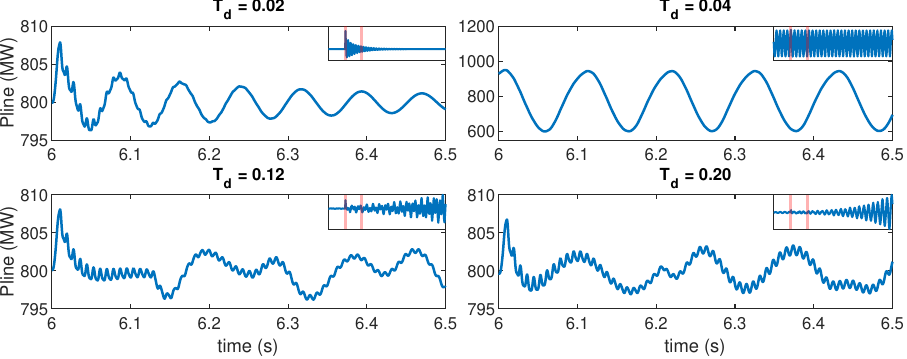}
    \caption{Active power of single delay system with different $T_d$}    \label{fig:SMIB_SingleDelay}
\end{figure}

\begin{figure}[htbp]
  \centering
  \begin{minipage}[b]{0.44\textwidth}
  \includegraphics[width=\textwidth]{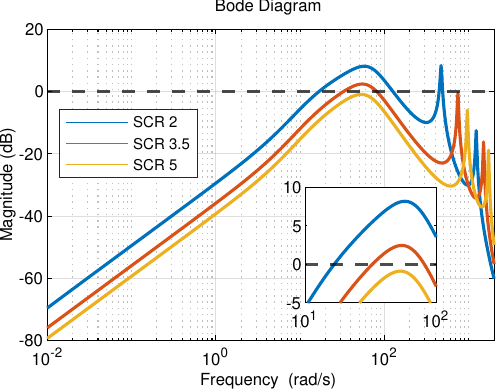}
    \caption{Frequency response for different SCRs.}
    \label{fig:SMIB_SingleDelaySCR}
  \end{minipage}
  \hspace{0.1in}
  \begin{minipage}[b]{0.44\textwidth}
  \includegraphics[width=\textwidth]{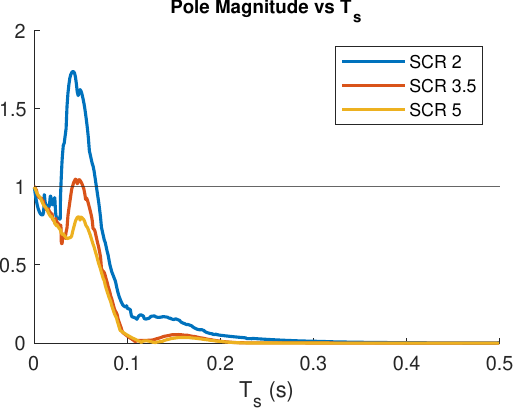}
    \caption{Magnitude of the unstable mode of the SMIB  system with different $T_s$}
    \label{fig:SMIB_SamplerSCR}
  \end{minipage}
\end{figure}

Assuming a decoupling between the active and reactive power controls enables the treatment of the two controls independently as two single-input single-output (SISO) systems. This allows the frequency response analysis of the SISO system using the open-loop characteristic, as discussed in Section \ref{sec:Analysis}. Let us now consider the effect of delay and sampling on the active power path of the GFL plant. The frequency response of the negative of the transfer function of the linearized system between the input to the delay element and the input to the sampler in the active power path (see Fig. \ref{fig:ControlArchitecture}) is shown for different values of delays in Fig. \ref{fig:SMIB_Bode}. The negation models the positive feedback that closes the loop and forms the closed-loop system.  The magnitude response shows that the system's gain is greater than unity between the two gain crossover frequencies at 32.5 rad/s and 81.2 rad/s. The closed-loop system becomes unstable if there is a phase crossover (i.e., crossing of the phase plot and the odd multiples of $180^{\circ}$ lines) in the range of [32.5, 81.2] rad/s. As discussed in Section \ref{sec:Analysis}, the phase crossovers shift with different values of delays and become denser for higher values of delays. The delay-free system has a phase crossover at 629 rad/s (not shown in the figure), which shifts to 84 rad/s with a 0.02 s delay, remaining outside the unstable range of [32.5, 81.2] rad/s. Increasing the delay to 0.04 s shifts the crossover into the unstable range at 60.8 rad/s, resulting in instability. At 0.12 s delay, although one crossover moves to 31 rad/s (just outside the unstable range), a second appears within it, maintaining instability. For larger delays (e.g., 0.2 s), the phase crossovers become increasingly dense, ensuring that at least one falls within the unstable range, thereby rendering the system unstable. Fig. \ref{fig:SMIB_SingleDelay} presents the response of the SMIB system with varying communication delay and sampling period set to 0. It demonstrates that the system's response is stable and poorly damped for a delay of 0.02 s, and is unstable for larger values of delay. Further, the frequency responses of the SMIB system for different SCRs are presented in Fig. \ref{fig:SMIB_SingleDelaySCR}. It can be observed that with a decrease in SCR, the range of frequency with positive gain increases, resulting in an increased region of instability with delays. 

\begin{figure}[htbp]
  \centering
     \includegraphics[width=0.8\textwidth]{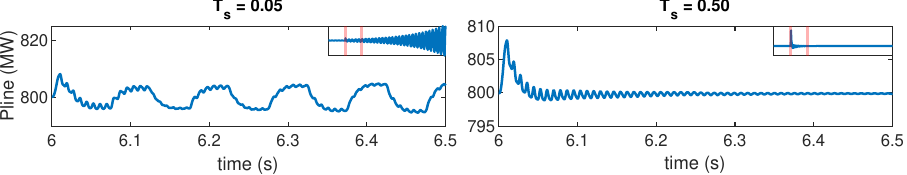}
    \caption{Active power of single sampler system with different $T_s$ for SCR 3.5.}
    \label{fig:SMIB_SingleSampler}
\end{figure}

\begin{figure}[htbp]
  \centering
    \includegraphics[width=0.8\textwidth]{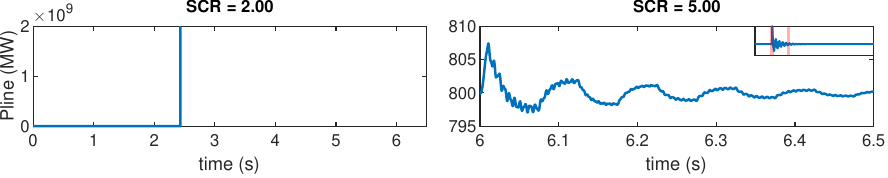}
    \caption{Active power of single sampler system with different SCRs for $T_s$ = 0.05s.}
    \label{fig:SMIB_SingleSamplerSCR}
\end{figure}

Similarly, to analyze the effect of sampling on stability, the variation of the magnitude of the discrete domain pole of the closed-loop system that is most influenced by sampling is presented in Fig. \ref{fig:SMIB_SamplerSCR}. The pole magnitude has a peak near the sampling period of 0.05 seconds. For a strong grid with SCR 5, the magnitude is always less than unity, and the system is stable for any value of sampling period. However, for weaker grids, with SCR 3.5 or 2, there exists a region where the pole magnitude exceeds unity, and the system is unstable for sampling periods corresponding to this region. Additionally, for all three SCR cases, the system is stable with large sampling periods. This is also evident from the low open-loop dc gain for the three cases in Fig. \ref{fig:SMIB_SingleDelaySCR}, as discussed in Section \ref{sec:Analysis}. Fig. \ref{fig:SMIB_SingleSampler} presents PSCAD simulation results showing that the system with SCR 3.5 is unstable for $T_s$ of 0.05 seconds but is stable for a large value of 0.5 seconds. Similarly, Fig. \ref{fig:SMIB_SingleSamplerSCR}, demonstrates that the system with SCR 2 is also unstable for $T_s$ of 0.05 seconds, but it becomes stable with SCR 5. This analysis provides insights on real events where oscillations or instability related to communication delay were observed following a disturbance that resulted in a decrease of SCR \cite{Chetan}. 

The above analysis provides insights into the individual effects of communication delay and sampling on small-signal stability of the SMIB system, but the discussed methods do not consider a system consisting of both communication delay and sampling period. Further, such observations cannot be made for systems with multiple IBRs because of the presence of multiple delays and samplers. This results in multi-input, multi-output (MIMO) open-loop systems with samplers operating at different rates, and the analysis does not hold. So, for the remainder of this section and the next Section, where the delays and samplers in both the active and reactive power paths are considered, the system behaviors are analyzed using time-domain simulations.

\subsection{Effect of Communication Delay}

\begin{figure}[htbp]
\centerline{\includegraphics[width=0.8\textwidth]{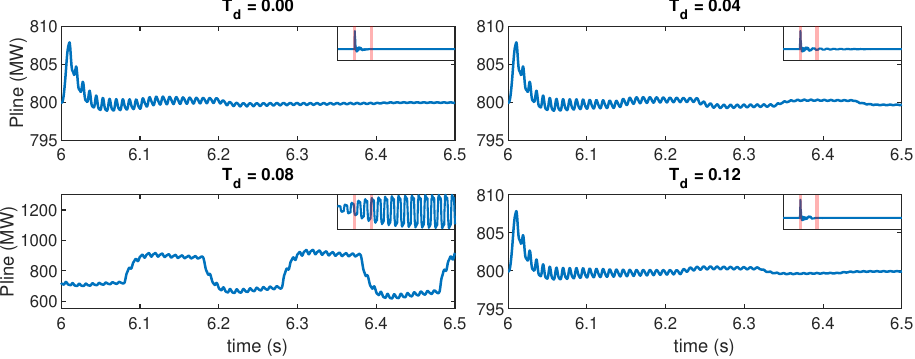}}
\caption{Active power of SMIB system with varying communication delay.}
\label{fig:DelaySMIB}
\vspace{-0.2in}
\end{figure}

With a fixed sampling period of 0.1 seconds, the time delay is varied to study the effect of communication delay between the plant-level control and the inverter-level control. Fig. \ref{fig:DelaySMIB} shows the response of the system with SCR 3.5 following the disturbance for different values of $T_d$, and the average values of the modal estimates obtained using the ring-down algorithms are presented in Table \ref{tab:SMIBDelayModes}. It can be observed that the damping of the dominant mode decreases from 38.00\% to 4.32\% as the delay is increased from 0 to 0.04 seconds. On further increasing the delay to 0.08 seconds, the system becomes small-signal unstable. The system returns to stability for a delay of 0.12 seconds. For the unstable case, the observed initial divergence followed by sustained oscillation could result from the unstable system response reaching the inverter limits.

\vspace{-0.15in}

\begin{table}[ht]
    \centering    
      \begin{minipage}[t]{0.48\textwidth}
      \centering
      \footnotesize
      \caption{Dominant Modes Estimates of SMIB system with Different Communication Delays.}
          \label{tab:SMIBDelayModes}
    \begin{tabular}{@{}c@{}c@{}c@{}c@{}}
        \toprule
        \textbf{\parbox{1.5cm}{\centering $T_d$ \\ (s)}} & \textbf{\parbox{1.5cm}{\centering Freq. \\ (Hz)}} & \textbf{\parbox{1.5cm}{\centering DR \\ (\%)}} & \textbf{\parbox{1.5cm}{\centering Energy \\ (\%)}} \\
        \midrule
        0.00 & 4.5 & 38.00 & 43.83 \\
        0.04 & 4.96 & 4.32 & 95.46 \\
        0.08 & 5.00 & -3.49 & 97.47 \\
        0.12 & 3.31 & 25.24 & 82.93 \\
        \bottomrule
    \end{tabular}
  \end{minipage}
    \hfill
  \begin{minipage}[t]{0.48\textwidth}
  \centering
  \footnotesize
  \caption{Dominant Modes Estimates of SMIB system with Different Sampling Periods.}
    \label{tab:SMIBSamplerModes}
    \begin{tabular}{@{}c@{}c@{}c@{}c@{}}
        \toprule
        \textbf{\parbox{1.5cm}{\centering $T_s$ \\ (s)}} & \textbf{\parbox{1.5cm}{\centering Freq. \\ (Hz)}} & \textbf{\parbox{1.5cm}{\centering DR \\ (\%)}} & \textbf{\parbox{1.5cm}{\centering Energy \\(\%)}} \\
        \midrule
        0.00 & 5.51 & -3.27 & 100 \\
        0.08 & 3.98 & 6.66 & 84.51 \\
        0.01 & 3.4 & 25.45 & 85.49 \\
        0.12 & 4.17 & -4.98 & 96.58 \\
        \bottomrule
    \end{tabular} 
  \end{minipage}
\end{table}

\subsection{Effect of Sampling Period}

\begin{figure}[htbp]
\centerline{\includegraphics[width=0.8\textwidth]{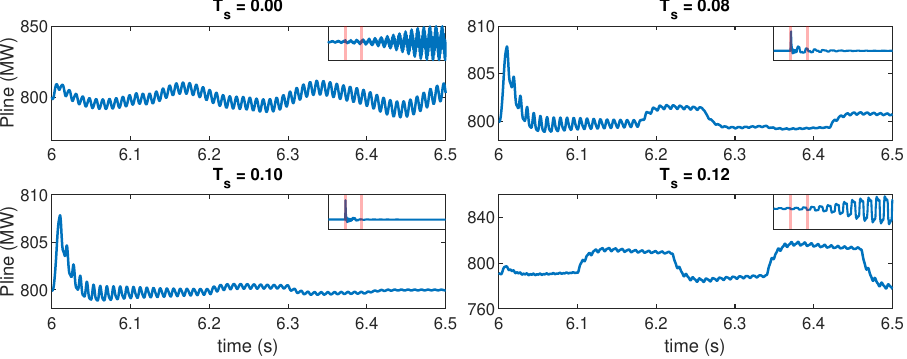}}
\caption{Active power of SMIB system with varying sampling time period.}
\label{fig:SamplerSMIB}
\vspace{-0.1in}
\end{figure}

With the communication delay set to 0.1 seconds, the sampling period is varied to study the effect of the sampling period between the plant level control and the inverter-level control. The responses of the system with SCR 3.5 following the disturbance for different sampling periods are presented in  Fig. \ref{fig:SamplerSMIB}, and the corresponding average values of estimated dominant modes are tabulated in Table \ref{tab:SMIBSamplerModes}. With no sampling of the control signal from the plant to the inverter control, the system is unstable, as seen for the case of $T_s=0$ seconds. The system becomes stable when the sampling period is increased to 0.08 seconds, and the dominant mode has a damping of 6.66\%. The damping significantly increases to 25.45\% as the sampling period is increased to 0.1 seconds. And, for the sampling period of 0.12 seconds, the system again becomes unstable. Such behavior, where the system exhibits multiple regions of stability and instability, is similar to that of the cases with k=0.5 and k=2, as discussed in Section \ref{sec:Analysis}. It is important to note that the delay-free system discussed in Fig. \ref{fig:SMIB_SingleSamplerSCR} has different regions of stability than this system with a delay of 0.1 seconds.

\subsection{Combined effect of Communication Delay and Sampling Time}

\begin{figure}[htbp]
\centerline{\includegraphics[width=0.8\textwidth]{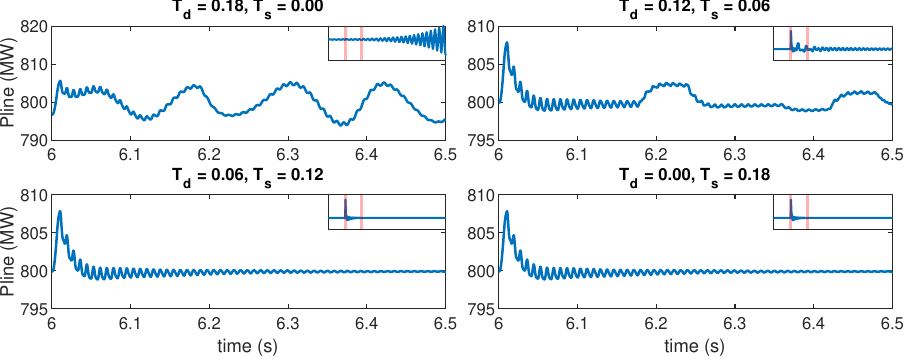}}
\caption{Different responses for same total delay of 0.18 s}
\label{fig:CombinedSMIB}
\vspace{-0.15in}
\end{figure}

A common practice found in the literature is to sum up the communication delay and the sampling period and model them together as a delay unit for stability analysis. In this section, we analyze the system response in different cases by varying the communication delay and sampling period such that their sum is 0.18 seconds. Fig. \ref{fig:CombinedSMIB} shows the response of the SMIB system to the disturbance under different cases. It can be observed that the responses in the four cases are significantly different. Table \ref{tab:ComparisonSMIB} presents the modal estimates obtained with the ring-down algorithms for these cases. The system's dominant mode is negatively damped with a damping of -2.55\% when $T_d = 0.18 s$ and $T_s = 0$. The system is poorly damped with the dominant mode's damping of 0.61\% when $T_d = 0.12s$ and $T_s = 0.06s$ while it is very well damped with the dominant mode's damping of 82.14\% and 81.31 \% when $T_d = 0.06 s, T_s = 0.12 s$, and $T_d = 0 s, T_s = 0.18 s$, respectively. The significant differences in these cases with the same sum of $T_d$ and $T_s$ highlight that bundling communication delay and sampling period together into a single delay entity for analysis can result in misleading stability information.

\begin{table}[ht]
\centering
    \begin{minipage}[t]{0.47\textwidth}
        \footnotesize
        \centering
        \caption{Dominant Mode Estimates of SMIB System with Same Total Sum of $T_d$ and $T_s$}
        \label{tab:ComparisonSMIB}
        \begin{tabular}{@{}c@{}c@{}c@{}c@{}c@{}}
        \toprule
        \textbf{\parbox{1.5cm}{\centering $T_d$ \\ (s)}} & \textbf{\parbox{1.5cm}{\centering $T_s$ \\ (s)}} & \textbf{\parbox{1.5cm}{\centering Freq. \\ (Hz)}} & \textbf{\parbox{1.5cm}{\centering DR \\ (\%)}} & \textbf{\parbox{1.5cm}{\centering Energy \\ (\%)}} \\
        \midrule
        0.18  & 0   & 7.98  & -2.55 & 100 \\
        0.12 & 0.06 & 8.33  & 0.61 & 100 \\
        0.06 & 0.12 & 2.64   & 82.14    & 100 \\
        0    & 0.18  & 2.74   & 81.31  & 100 \\
        \bottomrule
        \end{tabular}
    \end{minipage}%
    \hspace{0.1in}
    \hfill
    \begin{minipage}[t]{0.47\textwidth}
        \footnotesize
        \centering
        \caption{Dominant Mode Estimates of SMIB Systems with Different SCRs}
        \label{tab:SCRSMIB}
        \begin{tabular}{@{}c@{}c@{}c@{}c@{}}
        \toprule
         \textbf{SCR} & \textbf{\parbox{1.5cm}{\centering Freq. \\ (Hz)}} & \textbf{\parbox{1.5cm}{\centering DR \\ (\%)}} & \textbf{\parbox{1.5cm}{\centering Energy \\ (\%)}} \\
        \midrule
        10 & 4.77  & 23.60  & 94.19 \\
        5  & 5.05  & 8.95   & 81.84 \\
        3.5  & 5.07 & 2.25  & 95.23  \\
        2  & 4.92 & -0.75 & 50.94 \\
        \bottomrule
        \end{tabular}
    \end{minipage}    
\end{table}

\subsection{Sensitivity to Grid Strength}

\begin{figure}[htbp]
\centerline{\includegraphics[width=0.8\textwidth]{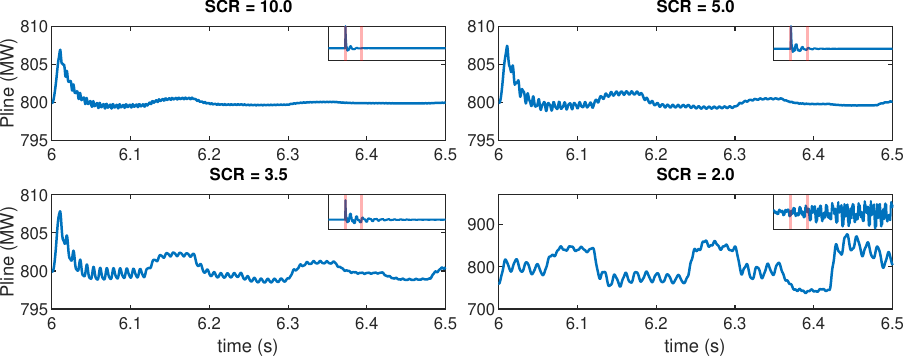}}
\caption{Effect on system with different SCRs}
\label{fig:SCRSMIB}
\vspace{-0.05in}
\end{figure}

The grid-following inverter's performance depends on the grid strength at the point of interconnection (POI). The grid strength at a point in a network is characterized by the SCR at the POI. To observe the effect of communication delay and sampling period on the SMIB systems with different SCRs, the grid impedance was varied, and the system was subjected to the same disturbance as discussed before. Fig. \ref{fig:SCRSMIB} shows the active power of the plant during the fault on systems with different SCRs. Both the communication delay and the sampling period were set to 0.06 seconds. The estimated modal properties are presented in Table \ref{tab:SCRSMIB}. The damping of the dominant mode of the system can be seen to decrease from a well-damped value of 23.60\% to -0.75\%, resulting in instability as the SCR is reduced from 10 to 2. For a given communication delay and sampling period, the system's small-signal stability becomes weaker with a decrease in SCR or grid strength. Conversely, the stability can be improved for a weak grid by adjusting the communication delay and the sampling period.

\subsection{Stability Regions}

\begin{figure}[htbp]
\centering
\begin{subfigure}[b]{0.3\linewidth}
\includegraphics[width=\linewidth]{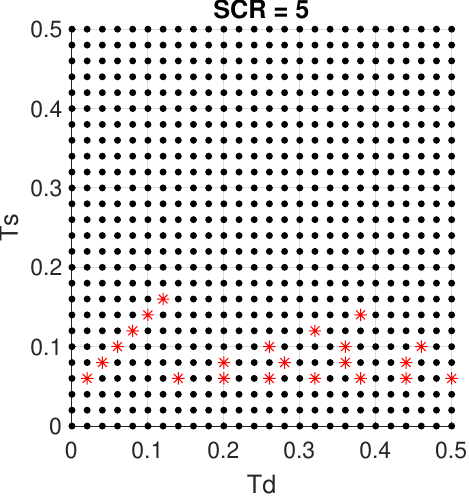}
\caption{Strong grid}
\label{fig:SMIBMatrixSCR10}
\end{subfigure}
\begin{subfigure}[b]{0.3\linewidth}
\includegraphics[width=\linewidth]{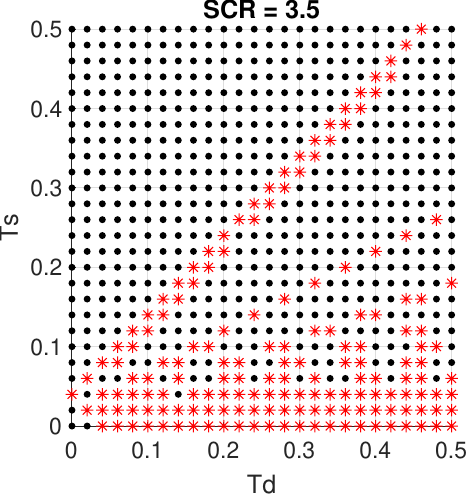}
\caption{Medium strength grid}
\label{fig:SMIBMatrixSCR3.5}
\end{subfigure}
\begin{subfigure}[b]{0.3\linewidth}
\includegraphics[width=\linewidth]{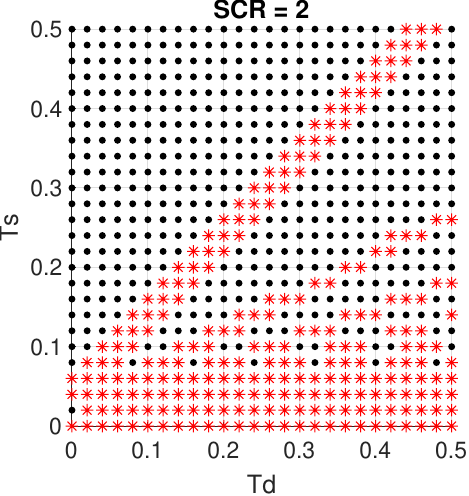}
\caption{Weak grid}
\label{fig:SMIBMatrixSCR2}
\end{subfigure}
\caption{Stability of SMIB system under with varying $T_d$ and $T_s$; Black: Small-signal stable, Red: Small-signal unstable}
\label{fig:SMIBMatrix}
\end{figure}

The small-signal stability of the SMIB system is affected by both sampling and communication delay between the plant-level control and the inverter-level control, and the system is more sensitive to their destabilizing effects for lower grid strength. Fig. \ref{fig:SMIBMatrix} shows the small-signal stability of the SMIB systems with strong, medium strength, and weak grids, with SCRs 5, 3.5 and 2, respectively, for different cases as both communication delay and sampling periods are varied from 0 to 0.5 seconds. For a very stiff grid with SCR 10, not shown in the figure, the system is stable for all tested values of $T_d$ and $T_s$. 

Fig. \ref{fig:SMIBMatrixSCR10} shows the regions of small-signal stability for the strong grid with SCR 5. It can be observed in Fig. \ref{fig:SMIB_SingleDelaySCR} that there is no gain crossover, and in Fig. \ref{fig:SMIB_SamplerSCR} that the dominant mode's magnitude is always less than unity for SCR 5. Both these observations indicate stable operation when the effect of either delay or sampling is considered individually. Such stable regions are observed for $T_s = 0$ and $T_d = 0$ in Fig. \ref{fig:SMIBMatrixSCR10}. However, small regions of instability are observed in Fig. \ref{fig:SMIBMatrixSCR10}, which result from the interaction of communication delay and sampling period and cannot be analyzed with the conventional methods.

\begin{figure}[htbp]
\centerline{\includegraphics[width=0.8\textwidth]{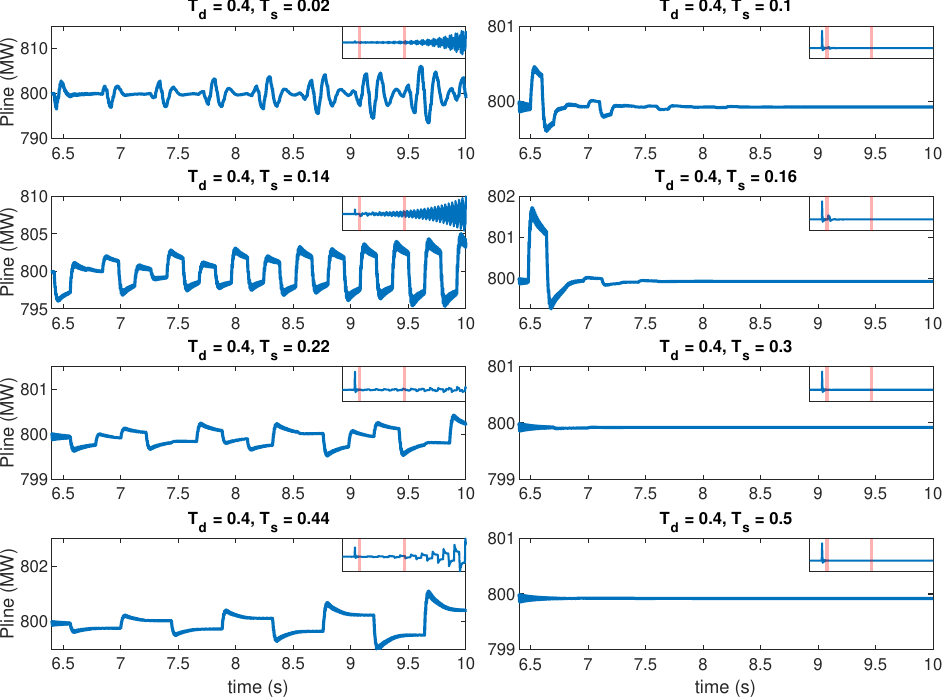}}
\caption{Active power of SMIB system with SCR 3.5, $T_d$ of 0.4 seconds and different $T_s$}
\label{fig:SMIB_Td0p4}
\vspace{-0.05in}
\end{figure}

For the medium strength grid with SCR 3.5, when $T_s$ is small, the system quickly becomes unstable on increasing the value of $T_d$, and remains unstable for higher $T_d$ values, as shown in Fig. \ref{fig:SMIBMatrixSCR3.5}. At low values of $T_s$, the destabilizing effect of communication delay, as discussed with Fig. \ref{fig:SMIB_Bode} for $T_s = 0$, is dominant and the system's stability is primarily affected by the communication delay. But, at higher values of $T_s$, the sampling effect helps to maintain stability, as shown in Fig. \ref{fig:SMIB_SamplerSCR} and even counters the negative impact of communication delay. The sampling effect can also cause instability, but it is in a small range of sampling periods, as seen for $T_d = 0$. Moreover, there are regions of instability even at high values of $T_s$, when large delays are considered. This is due to the combined influence of communication delay and sampling. Fig. \ref{fig:SMIB_Td0p4}, shows the response of the system post disturbance for $T_d = 0.4$ seconds and various $T_s$. For a low $T_s$ of 0.02 seconds, due to the destabilizing effect of the large delay, the system is unstable and has two negatively damped modes with frequencies of 8.33 Hz and 6.14 Hz as obtained with the ring-down algorithms. The interaction between the sampling and communication delay results in stable and very well-damped responses when $T_s$ is 0.1, 0.16, and 0.3 seconds, while the interaction results in unstable systems for $T_s$ values of 0.14, 0.22, and 0.44 seconds. The growing oscillations in these unstable cases are non-sinusoidal and can be observed to have frequencies of approximately 3.57 Hz (i.e. 0.28 s) for $T_s$ of 0.14 seconds, 0.45 Hz (i.e. 2.21 s) for $T_s$ of 0.22 seconds, and 1.14 Hz (i.e. 0.88 s) for $T_s$ of 0.44 seconds. When $T_s$ is 0.5s, the effect of the large sampling period makes the system stable. A thorough investigation of such interactions between communication delay and sampling, which can result in multiple regions of stability, is necessary to accurately assess the small-signal stability of IBR-rich power systems. The results also highlight the non-sinusoidal oscillations that can be observed in power systems due to the effect of communication delays and sampling in IBR plants.

Similarly, it can be observed in Fig. \ref{fig:SMIBMatrixSCR2} that for the weak grid with SCR 2, the regions are similar to those with SCR 3.5, but with greater regions of instability. The larger range of frequencies with gains greater than 0 dB in  Fig. \ref{fig:SMIB_SingleDelaySCR}, and the larger range of sampling period with pole magnitude greater than unity in Fig. \ref{fig:SMIB_SamplerSCR}, for SCR 2 indicate greater susceptibility of the weak grid to the destabilizing effects of communication delay and sampling period compared to strong grid and medium strength grids.

The vulnerability of the system stability with improper adjustment of communication delay and sampling period, and their significantly different impacts on stability, are evident with the behavior seen in Fig. \ref{fig:SMIBMatrix}. 

\section{Impact on Two Area Kundur System}
\label{sec:Kundur}

\begin{figure}[htbp]
\centerline{\includegraphics[width=0.6\textwidth]{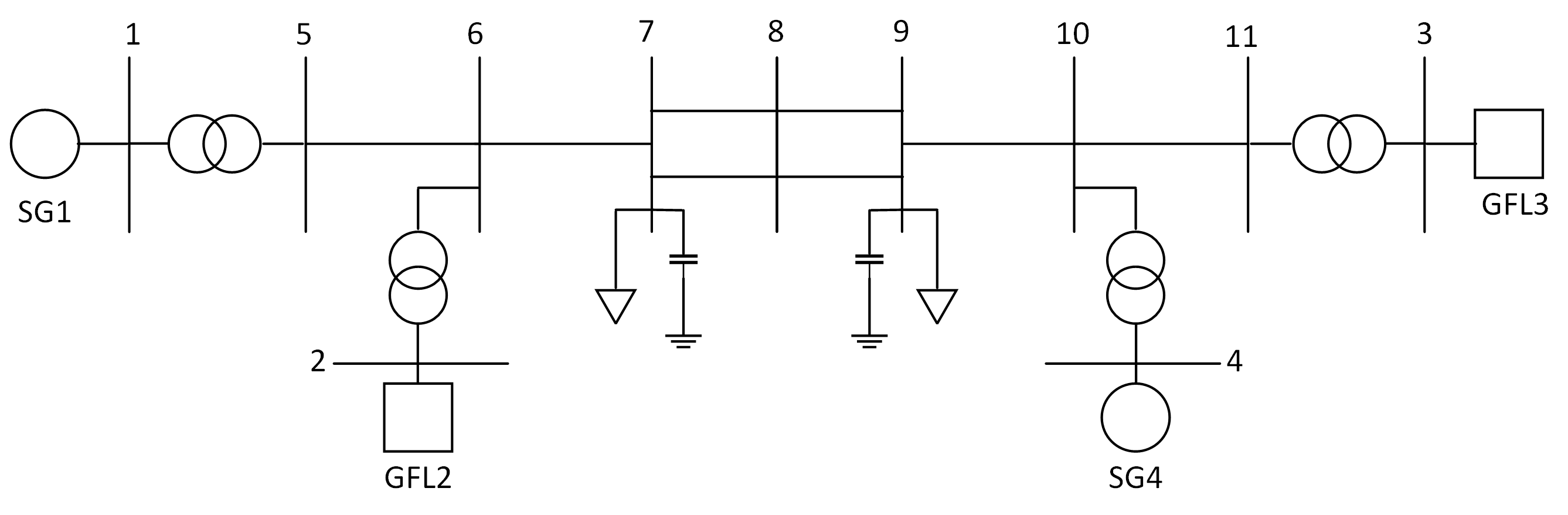}}
\caption{Modified two-area Kundur System}
\label{fig:Kundur}
\vspace{-0.1in}
\end{figure}

The two-area Kundur system presented in \cite{kundur1994power} is modified to contain two GFL power plants, one in each area, by replacing the synchronous generators in Bus 2 and Bus 3 as shown in Fig. \ref{fig:Kundur}. This system is used to study the effect of communication delay and sampling in GFL plants, considering their interaction with other dynamic components.

\begin{figure}[htbp]
\centerline{\includegraphics[width=0.45\textwidth]{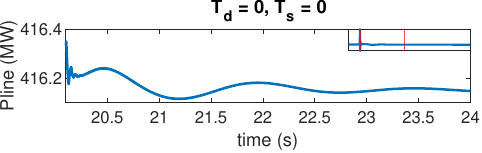}}
\caption{Tie-line flow with no delays and samplings}
\label{fig:Kundur00}
\vspace{-0.25in}
\end{figure}

The plant at Bus 2 is set to generate 700 MW of active and 215 MVAR of reactive power, while the plant at Bus 3 is set to generate 719 MW and 175 MVAR of active and reactive power, respectively. The loads are modeled as constant impedances and the synchronous generators are modeled using the PSCAD synchronous machine model equipped with SEXS\_PTI excitors and TGOV1 governors. The system is perturbed by a disturbance in the form of a pulse of magnitude of 0.01 pu and a duration of 0.01 seconds applied to the active power set point of the GFL plant at Bus 2. The main parameters of both inverters are the same and are presented in Table \ref{tab:GFLPara}. Fig. \ref{fig:Kundur00} shows the tie-line flow from Bus 7 to 8 when $T_d$ and $T_s$ of both the GFL plants are set to 0 seconds. At the steady state, 405 MW of active power is flowing through the tie-line and the system has a well-damped response to the disturbance.

\subsection{Effect of Individual Plant}

\begin{figure}[htbp]
  \centering
  \begin{subfigure}[b]{0.4\textwidth}
    \includegraphics[width=0.8\textwidth]{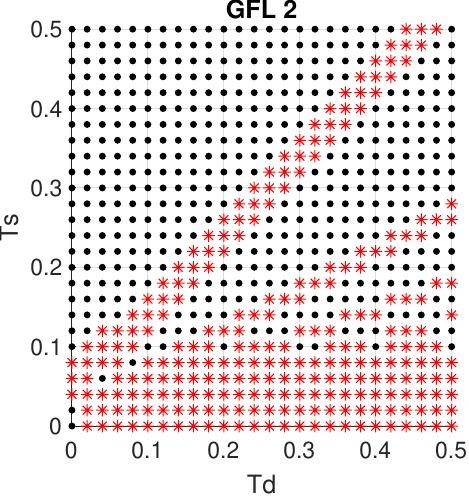}
    \caption{}
    \label{fig:KundurMatrix_GFL2}
  \end{subfigure}
  \hspace{0.1in}
  \begin{subfigure}[b]{0.4\textwidth}
  \includegraphics[width=0.8\textwidth]{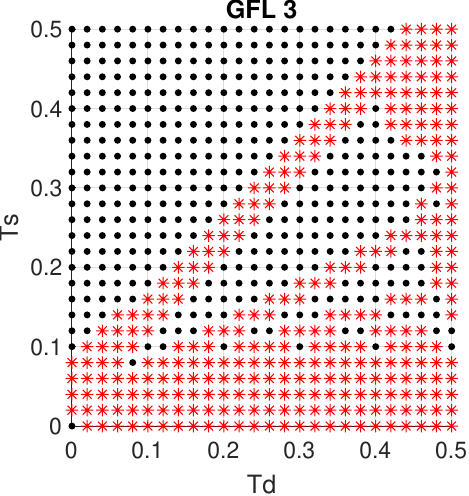}
    \caption{}
    \label{fig:KundurMatrix_GFL3}
  \end{subfigure}
  \caption{Stability of the Kundur System under different Td and Ts of (a) GFL 2 (b) GFL 3. Black dot: Small-signal stable, Red star: Small-signal unstable}
\end{figure}

\begin{figure}[htbp]
  \centering
     \includegraphics[width=0.8\textwidth]{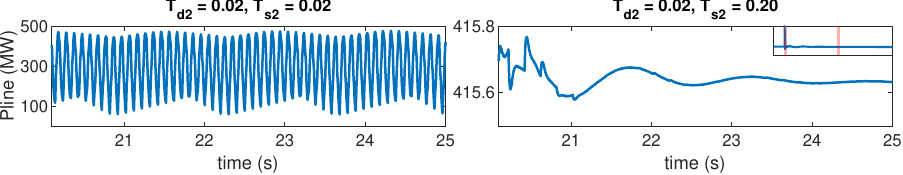}
    \caption{Tie-line flow for different $T_s$ with $T_d$ = 0.02 for GFL 2.}
    \label{fig:KundurGFL2}
\end{figure}

\begin{figure}[htbp]
  \centering
    \includegraphics[width=0.8\textwidth]{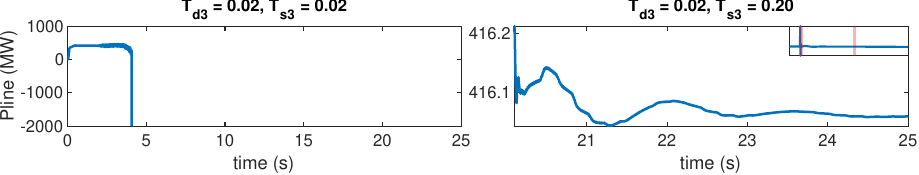}
    \caption{Tie-line flow for different $T_s$ with $T_d$ = 0.02 for GFL 3.}
    \label{fig:KundurGFL3}
  \end{figure}

The communication delay and the sampling period of the control signals from the plant-level control of each GFL plant were varied while keeping other parameters the same as the base case. Fig. \ref{fig:KundurMatrix_GFL2} shows the system's stability as Td and Ts of GFL 2 were varied. It can be observed that the regions of stability are similar to the case of low SCR in Fig. \ref{fig:SMIBMatrixSCR2}. Increasing communication delay in general compromises the stability of the system; however, the stability of the system can be achieved by using larger sampling periods. Such behavior is demonstrated in Fig. \ref{fig:KundurGFL2}, which shows that the system is unstable for a lower sampling time of 0.02 seconds but becomes well damped when the sampling time is increased to 0.2 seconds. Similar observations can be made for GFL 3 from Fig. \ref{fig:KundurMatrix_GFL3}. Further, Fig. \ref{fig:KundurGFL3} demonstrates the regain of stability with a higher sampling period for GFL 3, similar to the case of GFL 2. Compared to GFL 2, GFL 3 has a greater region of instability at higher values of Td, which can be attributed to the fact that the grid strength at Bus 3 is lower than that at Bus 2. 

\vspace{-0.1in}

\subsection{Effect of Multiple Plants}

\begin{figure}[htbp]
  \centering
  \begin{subfigure}[b]{0.4\textwidth}
  \includegraphics[width=0.8\textwidth]{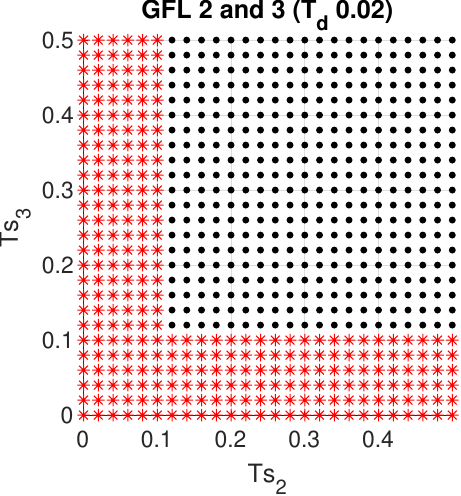}
    \caption{}
    \label{fig:KundurMatrixTd0p02}
  \end{subfigure}
  \hspace{0.1in}
  \begin{subfigure}[b]{0.4\textwidth}
  \includegraphics[width=0.8\textwidth]{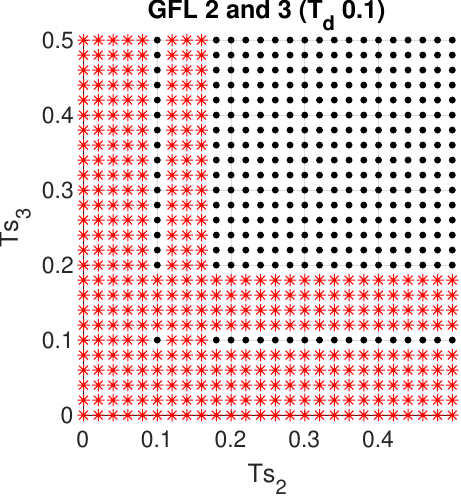}
    \caption{}
    \label{fig:KundurMatrixTd0p1}
  \end{subfigure}
  \caption{Stability of the Kundur System under with different samling periods (a) $T_d = 0.02 s$. (b) $T_d = 0.10 s$ Black dot: Small-signal stable, Red star: Small-signal unstable}
\end{figure}

In order to study the effect of sampling on multiple plants, the communication delay in both the GFL plants was fixed and the sampling period was varied. Fig. \ref{fig:KundurMatrixTd0p02} shows the region of stability when the communication delay was set to 0.02 seconds. The detrimental effect of communication delay makes the system unstable for small sampling periods of the GFL plants. However, when the sampling periods of both the plants are large enough to counter the impact of delays, the system becomes stable. Similarly, Fig. \ref{fig:KundurMatrixTd0p1} shows the stability plot for the case when communication delay is 0.1 seconds in both inverter plants. It can be noticed that to counter the destabilizing effect of this high delay, higher sampling periods are required in the GFL plants. Further, in case of high delay, multiple stability regions with small regions of stability can be noticed inside regions of instability. However, operation in the larger stability region i.e. with high sampling periods, is more conducive to system stability.

\section{Control Measures for Impact Mitigation}
\label{sec:Mitigation}

This section presents control strategies to improve the stability of IBR-rich systems, considering the effect of communication delay and sampling period. Fig. \ref{fig:SMIB_Bode} shows that the magnitude response of the SMIB system crosses the 0 dB line for SCR values of 3.5 and 2. Such gain crossovers create an increased chance of instability with communication delay as explained in Section \ref{sec:Analysis}. So, the stability of the system can be improved by adjusting the controls to avoid such gain crossovers. Further, Section \ref{sec:Analysis} also showed that for systems with open-loop dc gain less than unity, large sampling periods result in stable operation. Accordingly, this section presents two control strategies aimed at improving the stability of GFL plants, one employed at the plant-level control and the other at the inverter-level control.

\subsection{Use of Plant Level Low Pass Filter}

\begin{figure}[htbp]
  \centering
  \begin{minipage}[b]{0.45\textwidth}
  \includegraphics[width=\textwidth]{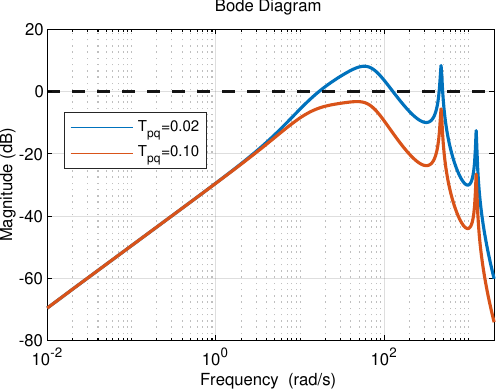}
    \caption{Frequency response with different time constants of plant-level filter.}
    \label{fig:TunedLPF}
  \end{minipage}
  \hspace{0.1in}
  \begin{minipage}[b]{0.45\textwidth}
  \centering
  \includegraphics[width=0.75\textwidth]{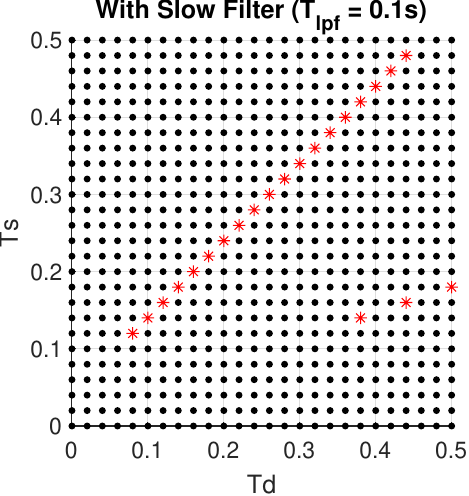}
    \caption{Stability of the system with the new plant-level filters.}
    \label{fig:TunedLPFMatrix}
  \end{minipage}
\end{figure}

Low-pass filters for the plant-level command signals, shown in Fig. \ref{fig:ControlArchitecture} with time constants $T_p$ and $T_q$, can be used to avoid or reduce the open-loop gains greater than unity. Fig. \ref{fig:TunedLPF} shows the magnitude responses of the active power control of the SMIB system for the case of SCR 2, with the plant-level low-pass filters with a time constant of 0.02 seconds, and with the time constant increased to 0.10 seconds. The lower bandwidth filter reduces the gain of the open-loop system at higher frequencies, thus avoiding the gain crossovers. Fig. \ref{fig:TunedLPFMatrix} shows that the new system with a filter time constant of 0.1 seconds has significantly higher regions of stability compared to Fig. \ref{fig:SMIBMatrixSCR2} with the time constant of 0.03 seconds. 

\subsection{Tuning Inverter Level Power Controller}

\begin{figure}[htbp]
  \centering
  \begin{minipage}[b]{0.45\textwidth}
  \includegraphics[width=\textwidth]{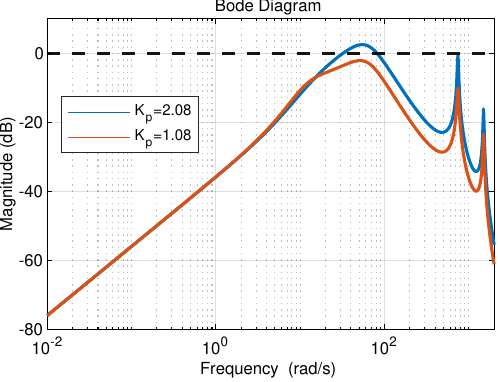}
    \caption{Frequency response with different time proportional gains of power controller.}
    \label{fig:TunedKp}
  \end{minipage}
  \hspace{0.1in}
  \begin{minipage}[b]{0.45\textwidth}
  \centering
  \includegraphics[width=0.75\textwidth]{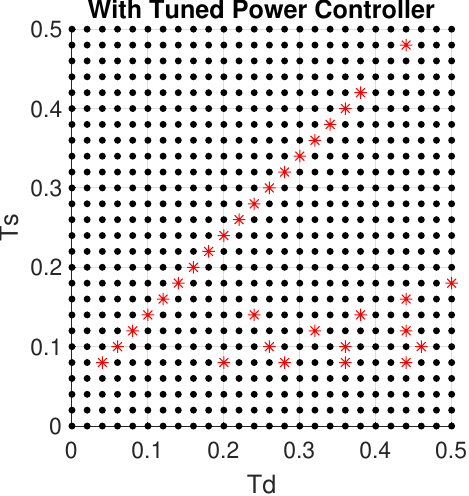}
    \caption{Stability of system with the new power controller.}
    \label{fig:TunedKpMatrix}
  \end{minipage}
\end{figure}

The controllers at the inverter level can also be tuned to avoid or reduce the open-loop gains greater than unity. Fig. \ref{fig:TunedKp} shows such a reduction in the system's gain to avoid 0 dB crossovers by changing the value of the proportional gain ($K_p$) of the power control loop from 2.08 to 1.08 for the SMIB system with SCR 3.5. Fig. \ref{fig:TunedKpMatrix} shows the stability of the SMIB system with the new power controller ($K_p = 1.08$). Compared with the base case, with $K_p=2.08$ in Fig. \ref{fig:SMIBMatrixSCR3.5},  a significant increase in the stability region can be observed with the new control. 

\subsection{Limitations of the control strategies}

The presented control strategies result in an increase of stable regions in the ($T_d, T_s$) space. Still, they do not guarantee the stability of the system in the entire region of interest, as both strategies assume decoupling of the active and reactive power control paths and do not completely consider the effect of sampling. The proposed strategies enhance stability by mitigating the destabilizing effect of communication delay, considering that large sampling periods lead to stable operations for systems with an open-loop DC gain below unity, as discussed in Section \ref{sec:Analysis}. However, as in the case with $k=0.5$ in Fig. \ref{fig:SamplerSMIB}, even such systems can become unstable for some low values of sampling periods. The destabilizing effects of sampling are over a much smaller range of values compared to those of communication delay, and hence, the proposed strategies can result in stability enhancement. A system can still exhibit regions of instability with the proposed controls, corresponding to the effects of sampling or the interaction between communication delay and sampling. This limitation of the analysis and the control strategies is reflected by the small, unstable regions with the tuned controls in Figs. \ref{fig:TunedLPFMatrix} and \ref{fig:TunedKpMatrix}. The analysis and stability assessment become more challenging for systems involving multiple IBR plants operating in proximity, and with different communication delays and sampling rates. The presence of multiple communication delays and samplers in a control loop results in a MIMO representation for the open-loop system, $G(s)H(s)$, as opposed to the SISO system discussed in Section \ref{sec:Analysis}. The frequency domain approach discussed in Section \ref{sec:Analysis:Delay} for the analysis of communication delay does not hold for MIMO systems, and the discrete analysis presented in Section \ref{sec:Analysis:Sampling} is not applicable to the study of systems with samplers operating at different rates. Hence, novel analytical approaches are required to accurately assess the small-signal stability of IBR-rich power systems, which can overcome such limitations and incorporate the coupled dynamics of communication delays and sampling in multiple IBR plants.

\section{Conclusion}
\label{sec:Conclusion}


This work has investigated the effects of the sampling period and communication delay between the plant-level control and inverter-level control of IBR power plants. Our findings demonstrate that the impact of sampling period and communication delay on a system's stability can be significantly different, and treating them as a single entity can lead to inaccurate assessments of system stability. The paper illustrates the complex interplay between communication delay and sampling period in time-delayed sampled systems, such as power systems with IBR plants, which results in multiple regions of stability. The paper also recommends preliminary control strategies to mitigate the detrimental effects of communication delays and sampling. The limitations of conventional analysis methods in capturing the impacts of delay and sampling on small-signal stability are discussed, and the importance of further research to accurately assess the small-signal stability of IBR-rich systems is highlighted. Future research can also explore the applicability of these findings to grid-forming inverters.

\vspace{-0.08in}
\section*{Acknowledgments}

The authors acknowledge the Power System Engineering Research Center (PSERC) and Réseau de Transport d'Électricité (RTE), France, for funding this research.

\bibliographystyle{elsarticle-num} 
\bibliography{main}
\end{document}